# Web Services Modeling and Composition Approach using Object-Oriented Petri Nets

Sofiane Chemaa[1], Raida Elmansouri[2] and Allaoua Chaoui[3]

[1] Department of Computer Science, University Mentouri, MISC Laboratory
Constantine, Algeria
*Chemaa@misc-umc.org*

[2] Department of Computer Science, University Mentouri, MISC Laboratory
Constantine, Algeria
*raidaelmansouri@yahoo.fr*

[3] Department of Computer Science, University Mentouri, MISC Laboratory
Constantine, Algeria
*a_chaoui2001@yahoo.com*

**Abstract**

Nowadays, with the emergence and the evolution of new technologies, such as e-business, a large number of companies are connected to Internet, and have proposed web services to trade. Web services as presented, are conceptually limited components to relatively simple functionalities. Generally, a single service does not satisfy the users needs that are more and more complex. Therefore, services must be made able to be composed to offer added value services. In this paper, a web services composition approach, modelled by Objects-Oriented Petri nets, is presented. In his context, an expressive algebra, which successfully solves the web services complex composition problem, is proposed. A java tool that allows automating this approach; based on a definite algebra and a G-nets meta-model, proposed by us, is developed.

.

## 1. Introduction

The emergence of the web services paradigm has marked a significant evolution in the history of the internet, which was intended to play the role of a means of exchanging data. With web services, Internet is converted to an easily integrable, loosely-coupled and self-describing software component platform [1]. Web services are to overcome the problems encountered by the companies in terms of interoperability, by implementing a SOA (Service Oriented Architecture) [2] based on a set of standards. The process of standardization affects three layers of the infrastructure based on Service Oriented Architecture: the communication protocol, which allows the structure of the exchanged messages between the services, the specification of the services interface description; and finally the publication specification, and the services localization.

The concept of web service basically refers to an application on internet. It is made available by a service provider and used by customers via standard internet protocols such as Universal Description, Discovery, and Integration (UDDI) [3], Web Service Description Language (WSDL) [4] and Simple Object Access Protocol (SOAP) [5].

Reutilization is one of the most important advantages of the web service paradigm. Web services, as they are presented, are conceptually limited to relatively simple functionalities, which are modeled by a set of operations. However, it is necessary to set up new applications by composing services to counter more complex requirements [6].

The web services composition is a natural evolution of this technology; it has a remarkable potential ability in enterprise application integration and business to business reorganization. Although solutions for description, publication, discovery, and interoperability of web services are provided by the recent technologies based on WSDL, UDDI, and SOAP, complex composition cannot be achieved. SOAP is a simple XML-based protocol to let applications exchange information over HTTP. WSDL is a general purpose XML language for describing what a Web service does, where it resides, and how to invoke it. The UDDI standard is a directory service that contains service publications and enables web-service clients to locate candidate services and discover their details. Nevertheless, the web services composition remains a very complex task, and requires formal techniques for its accomplishment.

In this article, we tackle the problem of web services composition using a formalism based on Petri-nets [7], called G-nets [8]. Comparing with the other approaches, this concept offers efficient and powerful mechanisms for modeling complex systems, which are not supported even by high level Petri-nets. In this context, a G-net based algebra, which succeeds in solving the web services composition problem, is proposed. In addition to a formalism allowing to describe the services behavior and structure, the proposed approach provides an operators representative set, presented via their syntax and their semantics. After which we present a Java tool, which enables to automate this approach on the basis of the use of a meta-model defined for G –nets, and all the formally defined operators. The use of a formal model allows the verification of properties and the detection of inconsistencies both within and between services.

This paper is divided into eight major sections. In addition to the introductory section, section two presents the web services modeling and their specifications, using G-Nets. Section three is devoted to algebra for composing web services and its G-Net based formal semantics. Section four, includes a meta-model for G-nets. Section five, presents an application permits to specify web services by G-nets, and automate the different composition rules using the programming language, JAVA. Section six, deals with the verification topic. Before concluding, we presented a brief overview of some related works. Finally, we discussed works in process, and some perspectives.

## 2. Modeling web services using G-nets

We briefly begin with an informal definition of the elementary Petri net [7]. Petri net is a means for modeling the behavior of Discrete Event Dynamic Systems. This is a directed bipartite graph with two types of nodes: places are represented by circles and transitions are represented by rectangles. A place can hold a finite number of tokens and this nonnegative integer is called the marc of the place. The arcs of the graph connect places and transitions in such a way that places can only be connected to transitions and vice versa. An arc is labeled with nonnegative integer called the weight. A Petri-net is a diagram with operational semantics, i.e. a behavior is associated with a diagram, which enables to describe the represented system dynamics.

G-Net is a Petri Net based framework introduced by [8]. It is utilized in the modular design and specification of complex and distributed information systems. A formalism that adopts object oriented structuring into Petri Nets broadly, is provided by this framework. Advantages from the formal treatment and the expressive comfort of Petri nets are intended to be taken, and benefits from object oriented approach (reusable software, extensible components, encapsulation, etc.), are planned to be gained. A system designed by the G-net framework is made up of a set of autonomous and loosely coupled modules called G-Nets. The encapsulation property is satisfied by a G-net i.e. a module cannot be accessed by another module, but through a G-net abstraction, which is a well-defined mechanism, like an object in the object oriented programming concept.

A G-net consists of two parts. The Generic Switch Place (GSP), which is a special place that represents the visible part of a G-net i.e. the interface between a G-net and other ones. Besides, the Internal Structure (IS), the G-net invisible part, which constitutes the internal realization of the designed system. Both, specifying notations of the IS and the Petri net are very close. For more details about the Petri nets and G-nets, the reader is referred to [9],[8] and [10].

Web services resemble distributed system, which is made up of a set of loosely coupled modules communicating through exchanging messages as a G-net system. Therefore, we model web services using G-Net, easily. We model each operation in the service by a method in the G-net and associate to each one a piece of Petri-Net in the IS of the G-Net. Consequently, we model the state of the service by the position of the tokens in the G-Net.

Next, some formal definitions about G-Net service and Web service are given.

***Definition 1. (G-Net Service)*** A G-net service is a G-Net $S(GSP, IS)$ where:

- $GSP(MS, AS)$ is a special place that represents the abstraction of the service where:
  - MS is a set of executable methods in the form of $< MtdName >< description >= \{ [P1: description; ......; Pn: description] (< InitPL >) (< GoalPls >)\}$ where $< MtdName >$ and $< description >$ are the name and the description of the method respectively. $< P1 : description;......; Pn : description >$ is a set of arguments for the method, $< InitPL >$ is the name of the initial place for the method and $< GoalPls >$ is(are) the name(s) of the goal place(s) for the method.
  - AS is a set of attributes in the form of $< attribute - name > = \{< type >\}$ where $< attribute\text{-}name >$ is the name of the attribute and $< type >$ is the type of the attribute.

- $IS(P, T, W, F, Trc, Tra, L)$ is the internal structure of the service, a modified predicate/transition net [11], where:
  - $P = NP \cup ISP \cup GP$ is a finite and non-empty set of places where NP is a set of normal places denoted by circles, ISP is the set of instantiated switch places denoted by ellipses used to interconnect G-Nets, GP is the set of goal places denoted by double

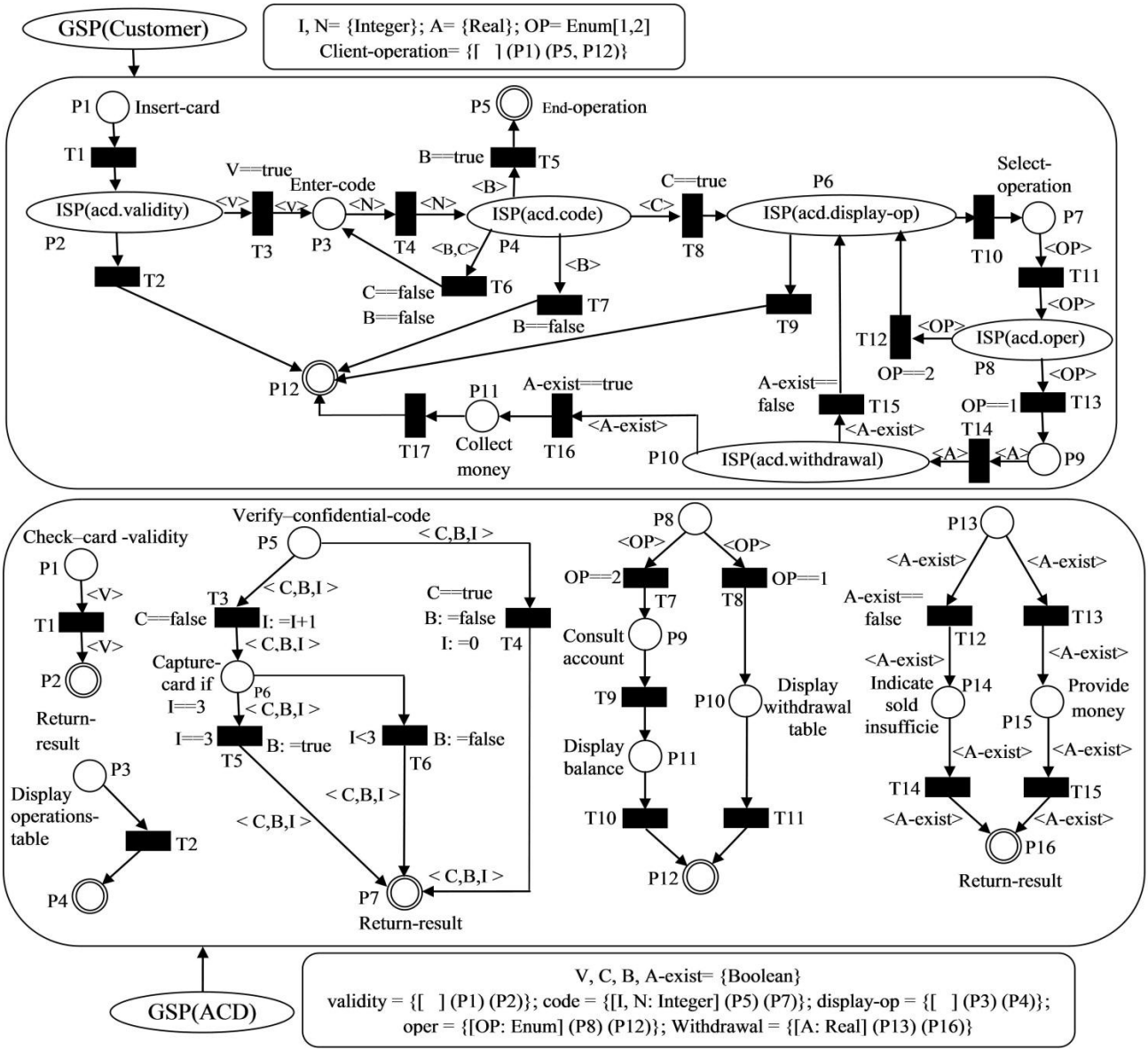

Fig. 1 Example of G-net Services

- circles used to represent final state of method's execution.
- T is a set of transitions.
- W is a set of directed arcs W ⊆ (P × T) ∪ (T × P) (the flow relation).
- F is an application that associates a description to certain elements of W.
- Trc is an application that associates a condition to certain transitions (called selectors of transition), this condition is a logical formula constructed from variables which appeared in the inscriptions of adjacent input-arcs.
- Tra is an application that associates an action to certain transitions, this action is a sequence of affectations of values to variables.
- L : P → O ∪ {τ} is a labeling function where O is a set of operation names and τ is a silent operation.

In figure 1, we propose an illustrative example of two services customer and automatic cash dispensers (ACD) represented by G-Nets. The service customer reproduces the behavior of the client, who wants to use the ACD that insures the credit consultation, and cash withdrawals operations. For this we must have a valid magnetic card and a confidential code. To accomplish each operation this card must be inserted into the machine which automatically verifies its validity. In the affirmation the customer then is invited to introduce his confidential code, which is composed of numbers, by using the machine keyboard. If the code is accepted the operations should be displayed on the screen by the ACD, therefore the desired operation is selected by the customer pressing the suitable icon. Two cases to consider:

- If it concerns an application for credit, the machine displays the account balance, whereupon it redisplays the operations table.
- If it concerns funds withdrawal application, the machine displays a window related to this operation, which allows the client to introduce the sum he wants to withdraw. After the validation, the validity of the sum is automatically checked by the machine. If the sum is not sufficient; the machine would make an indication on the screen, it redisplays the operations table. In the case of the availability of funds, it will distribute the banknotes representing the inserted sum.

However, if the confidential code is erroneous after three successive attempts, the magnetic card will be captured by the ACD for security purposes, because it may be that the user of the card is not its owner. Its recuperation depends on a special processing that is not tackled in this example.

It should be noted that the customer may change his opinion, and interrupt any operation at any time before the validation.

The ISP notation serves as the primary mechanism for specifying the interconnection between different G-Nets. In the example of Figure 1, we integrate the ISP of ACD with its methods validity, code, display-op, oper and withdrawal in the IS of Customer to specify a client-server relation.

*The significance of the using attributes:*

- V: indicates either the card is valid or not.
- C: indicates either the confidential code is valid or not.
- B: indicates either the card is blocked or not.
- A-exist: indicates either the sum to withdraw is available in the account or not.
- I: represents the number of successive errors during the introduction of the confidential code (at each incorrect attempt the machine increment by 1 the number I, if the introduced code is accepted the machine affect the value 0 to I).
- N: represents the confidential code.
- A: represents the sum to withdraw.

- OP: represents the type of the operation (withdraw if OP==1 or credit consultation if OP==2).

*Definition 2. (Web Service)* A Web service is a tuple [12] $S = (NameS, Desc, Loc, URL, CS, SGN)$ where:
- NameS is the name of the service used as its unique identifier.
- Desc is the description of the provided service. It summarizes what functionalities the service offers.
- Loc is the server in which the service is located.
- URL is the invocation of the Web service.
- CS is a set of the component services of the Web service, if CS = {NameS} then S is a basic service, otherwise S is a Composite service.
- $SGN = (GSP, IS)$ is the G-Net modeling the dynamic behavior of the service.

The concept of G-Net service and Web service being presented, we show in the next section how Web services can be incrementally composed. We recall that we use G-Nets as a means to offer a flexible and powerful algebra.

## 3. Web Services Composition

Generally, a single web service does not satisfy the users' needs, which are increasingly complex. By definition, web services as presented, are components, conceptually limited, with relatively simple functionalities, which are modeled by a set of operations [13]. The cooperation of two or more different web services leads to a novel task achievement. Consequently, a new value to the services collection is added, giving the example of the collaboration of a Hotel Booking Service and a Web Mapping Service such as Google Maps API for the clients' guidance. Hence, a complex Web service performing the original tasks in addition to a new one is the result of the cooperation of these services.

In this section we present an algebra that combines existing Web services for building more complex ones. Sequence, Parallel, Alternative, Iteration, and Arbitrary Sequence as basic constructs will be taken. Moreover, four more developed constructs which are Discriminator, selection, refinement and replace are defined. After that, each operator formal semantics in terms of G-nets, after its informal definition provision, is given.

3.1 Composition constructs

The BNF-like notation below describes a grammar defining the set of services that can be generated using algebra's operators.

S::= ε | X | S▶S | S◀▶S | ↺ S | S ⇔ S | S//S | (S ▫ S)
≫ S | [S..|..S] | $Ref(S, a, A)$ | $Rep(S, S, S)$ where:

- ε is the Zero Service (or empty service), i.e a service which performs no operation.
- X is a constant service. It consists of a service performing operation that cannot be split into sub-operations. This service is called Atomic.
- $S_1 ▶ S_2$ represents a composite service that performs one service immediately followed by another, i.e ▶ is a sequence operator.
- $S_1 ◀▶ S_2$ represents a composite service that can reproduce either the behavior of $S_1$ or $S_2$, i.e ◀▶ is an alternative ( or a Mutual Exclusion) operator.
- ↺ S represents a composite service where one service is successively executed multiple times, i.e ↺ is an iteration operator.
- $S_1 ⇔ S_2$ is a composite service that performs any arbitrary sequence of the services $S_1$ and $S_2$, i.e ⇔ is an unordered sequence operator.
- $S_1 //S_2$ represents a composite service which performs the two services $S_1$ and $S_2$ at the same time and independently. The resulting service waits until the end of execution of $S_1$ and $S_2$, i.e // is a parallel operator.
- $(S_1 \; ▫ \; S_2 \; ▫ \; ... \; ▫ \; S_{n-1}) ≫ S_n$ is a composite service that waits for the execution of one service (among the n-1 services), before activating the subsequent service $S_n$, i.e (.▫.) ≫. is a discriminator operator. Note that the n-1 first services are performed in parallel and without communication.
- $[S_1 | S_2 | ... | S_n]$ represents a composite service that selects and executes another service among n services available that perform the same task, i.e [..|..] is an operator of selection.
- $Ref(S, a, A)$ represents a service which behaves as S except for the operation labeled by "a", which is replaced by "A". This latter is a modified predicate/transition net, i.e $Ref( )$ is a refinement operator.
- $Rep(S, S_1, S_2)$ represents a composite service similar to S except for the service component $S_1$, which is replaced by $S_2$ not empty, i.e $Rep( )$ is a replace operator.

The proposed algebra verifies the closure property. This latter ensures that the product of any operation on services is itself a service to which we can apply algebra operators. We are thus able to build more complex services by aggregating and reusing existing services through declarative expressions of service algebra.

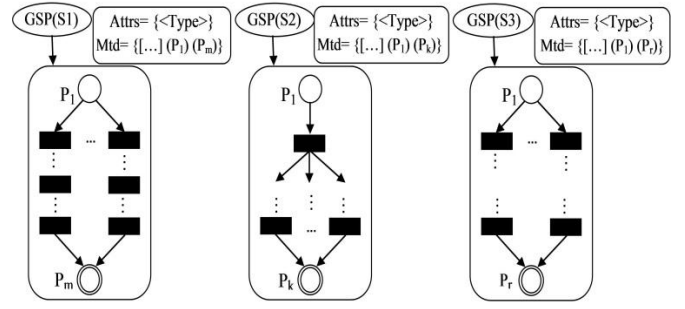

Fig. 2 Services S1,S2 and S3

3.2 Formal Semantics

In this section, we give a formal definition, in term of G-Nets, of the composition operators. Let $S_i = (NameS_i, Desc_i, Loc_i, URL_i, CS_i, SGN_i)$ with $SGN_i = (P_i, T_i, W_i, F_i, Trc_i, Tra_i, L_i)$ for i = 1,..,n be n Web Services such that $P_i \cap P_j = \emptyset$ and $T_i \cap T_j = \emptyset$ for $i \neq j$.

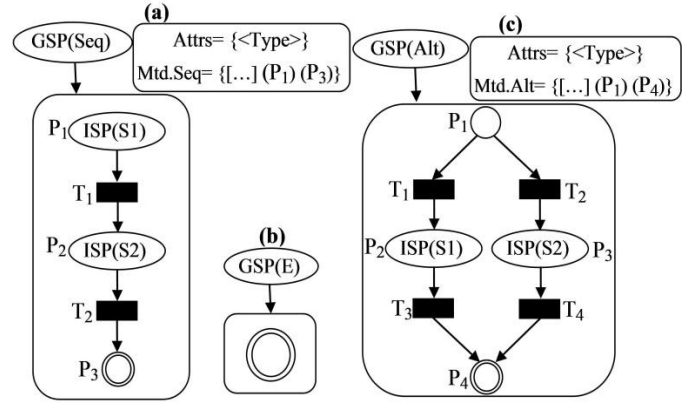

Fig. 3 Sequence service (a), Empty service (b) and Choice service (c)

**Empty Service.** The empty service ε is a service that performs no operation. It is used for technical and theoretical reasons.

***Definition 3***. The Empty service is defined as ε = $(NameS, Desc, Loc, URL, CS, SGN)$ where:
- $NameS$ = Empty.
- $Desc$ = "Empty Web Service".
- $Loc$ = Null, stating that there is no server for the service.
- $URL$ = Null, stating that there is no invocation for the service.
- $CS$ = {Empty}  ● $SGN = (GSP, IS)$ where :

  - GSP (MS, AS) where MS = ∅, AS = ∅.
  - $IS = (p, \emptyset, \emptyset, \emptyset, \emptyset, \emptyset, \emptyset)$.

In Figure 3(b), we show the graphic representation of the empty service (ε) in terms of G-Net.

In the following definitions, NameS is the name of the new service; Desc is the description of the new service; Loc is the location of the new service; URL is the invocation of the new service.

**Sequence.** The sequence operator allows the construction of a service composed of two services executed one after the other. This construction is used when a service should wait the execution result of another one before starting its execution. For example when subscribing to a forum, the service *Registration* is executed before the service *Confirmation*.

**Definition 4.** The service $S_1 \blacktriangleright S_2$ is defined as $S_1 \blacktriangleright S_2 = (NameS, Desc, Loc, URL, CS, SGN)$ where:
- $CS = CS_1 \cup CS_2$
- $SGN = (GSP, IS)$ where :

  - GSP $= (MS, AS)$ where
    $MS = Mtd.Seq\{[...](p_1)(P_3)\}, AS = [...]$.
  - $IS = (P, T, W, F, Trc, Tra, L)$ where $P = \{p_1, p_2, p_3\}, T = \{t_1, t_2\}, W = \{(p_1, t_1), (t_1, p_2), (p_2, t_2), (t_2, p_3)\}, F = \{...\}, Trc = \{...\}, Tra = \{...\}, L = \{(p_1, ISP(S_1)), (p_2, ISP(S_2)), (p_3, goal)\}$.

Given the two services $S_1$ and $S_2$ shown in Figure 2, the composite service $S_1 \blacktriangleright S_2$ is represented by the G-Net shown in Figure 3(a).

**Alternative.** Given two services $S_1$ and $S_2$, the alternative operator reproduces either the behavior of $S_1$ or $S_2$, but not both. For example the service Identification is followed either by the service *Allow-access* or the service *Deny-access*.

**Definition 5.** The service $S_1 \blacktriangleleft \blacktriangleright S_2$ is defined as $S_1 \blacktriangleleft \blacktriangleright S_2 = (NameS, Desc, Loc, URL, CS, SGN)$ where:
- $CS = CS_1 \cup CS_2$
- $SGN = (GSP, IS)$ where :

  - GSP $= (MS, AS)$ where
    $MS = Mtd.Alt\{[...](p_1)(P_4)\}, AS = [...]$.
  - $IS = (P, T, W, F, Trc, Tra, L)$ where $P = \{p_1, p_2, p_3, p_4\}, T = \{t_1, t_2, t_3, t_4\}, W = \{(p_1, t_1), (t_1, p_2), (p_2, t_3), (t_3, p_4), (p_1, t_2), (t_2, p_3), (p_3, t_4), (t_4, p_4)\}, F = \{...\}, Trc = \{...\}, Tra = \{...\}, L = \{(p_1, \tau), (p_2, ISP(S_1)), (p_3, ISP(S_2)), (p_4, goal)\}$.

Given the two services $S_1$ and $S_2$ shown in Figure 2, the composite service $S_1 \blacktriangleleft \blacktriangleright S_2$ is represented by the G-Net shown in Figure 3(c).

**Iteration.** The iteration operator allows the service S to be performed a certain number of times in a row. An example of use of this construct is when a customer orders a good a certain number of times from a service.

**Definition 6.** The service $\circlearrowleft S_1$ is defined as $\circlearrowleft S_1 = (NameS, Desc, Loc, URL, CS, SGN)$ where:

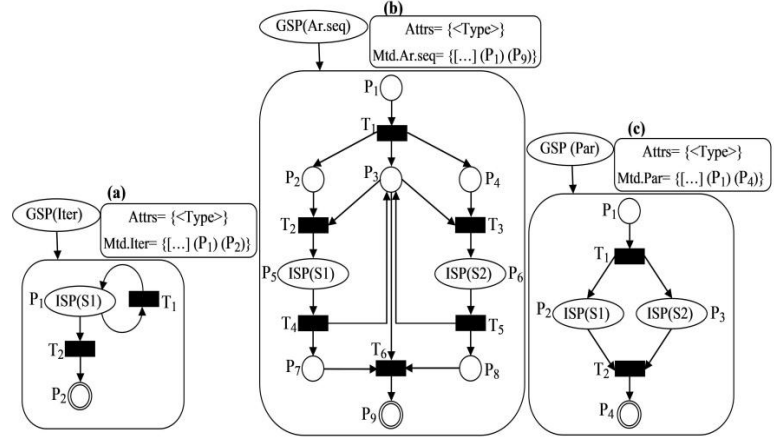

Fig. 4 Iteration service (a), Arbitrary sequence service (b) and Parallel service (c)

- $CS = CS_1$
- $SGN = (GSP, IS)$ where :

  - GSP $= (MS, AS)$ where
    $MS = Mtd.Iter\{[...](p_1)(P_2)\}, AS = [...]$.
  - $IS = (P, T, W, F, Trc, Tra, L)$ where $P = \{p_1, p_2\}$ $T = \{t_1, t_2\}, W = \{(p_1, t_1), (t_1, p_1), (p_1, t_2), (t_2, p_2)\}, F = \{...\}, Trc = \{...\}, Tra = \{...\}, L = \{(p_1, ISP(S_1)), (p_2, goal)\}$.

If we consider the service $S_1$ shown in Figure 2, the composite service $\circlearrowleft S_1$ is represented by the G-Net shown in Figure 4(a).

**Arbitrary Sequence.** The arbitrary sequence operator specifies the execution of two services that must not be executed concurrently. This construct is useful when there are no benefits to execute services in parallel. For example when there is no deadline to accomplish the global task and the parallelism generates additional costs.

**Definition 7.** The service $S_1 \Leftrightarrow S_2$ is defined as $S_1 \Leftrightarrow S_2 = (NameS, Desc, Loc, URL, CS, SGN)$ where:
- $CS = CS_1 \cup CS_2$
- $SGN = (GSP, IS)$ where :

  - GSP $= (MS, AS)$ where
    $MS = Mtd.Ar.seq\{[...](p_1)(P_9)\}, AS = [...]$.
  - $IS = (P, T, W, F, Trc, Tra, L)$ where $P = \{p_1, p_2, p_3, p_4, p_5, p_6, p_7, p_8, p_9\}, T = \{t_1, t_2, t_3, t_4\}, W = \{(p_1, t_1), (t_1, p_2), (t_1, p_3), (t_1, p_4), (p_2, t_2), (t_2, p_5), (p_5, t_4), (t_4, p_7), (t_4, p_3), (p_7, t_6), (t_6, p_9), (P_3, t_2), (p_3, t_3), (p_3, t_6), (p_4, t_3), (t_3, p_6), (p_6, t_5), (t_5, p_3), (t_5, p_8), (P_8, t_6)\}, F = \{...\}, Trc = \{...\}, Tra = \{...\}, L = \{(p_1, \tau), (p_2, \tau), (p_3, \tau), (p_4, \tau), (p_7, \tau), (p_8, \tau), (p_5, ISP(S_1)), (p_6, ISP(S_2)), (p_9, goal)\}$.

Given the two services $S_1$ and $S_2$ shown in Figure 2, the composite service $S_1 \Leftrightarrow S_2$ is represented by the G-Net shown in Figure 4(b).

**Parallel.** Given two services $S_1$ and $S_2$, the parallel operator builds a composite service performing the two services ($S_1$ and $S_2$) in parallel and without interaction between them. The accomplishment of the resulting service is achieved when the two services are completed. This construct is useful when a service executes multiple atomic services completely independent.

***Definition 8.*** The service $S_1 // S_2$ is defined as $S_1 // S_2 = (NameS, Desc, Loc, URL, CS, SGN)$ where:
- $CS = CS_1 \cup CS_2$.  • $SGN = (GSP, IS)$ where:

  - $GSP = (MS, AS)$ where
    $MS = \text{Mtd. Par}\{[...](p_1)(P_4)\}, AS = [...]$.
  - $IS = (P, T, W, F, Trc, Tra, L)$ where $P = \{p_1, p_2, p_3, p_4\}, T = \{t_1, t_2\}, W = \{(p_1, t_1), (t_1, p_2), (t_1, p_3), (p_2, t_2), (p_3, t_2), (t_2, p_4)\}, F = \{...\}, Trc = \{...\},$
    $Tra = \{...\}, \quad L = \{(p_1, \tau), (p_2, ISP(S_1)) (p_3, ISP(S_2)), (p_4, goal)\}$.

Given the two services $S_1$ and $S_2$ shown in Figure 2, the composite service $S_1 // S_2$ is represented by the G-Net shown in Figure 4(c).

***Definition 9.*** The service $(S_1 \square S_2 \square ... \square S_{n-1}) \gg S_n$ is defined as $(S_1 \square S_2 \square ... \square S_{n-1}) \gg S_n = (NameS, Desc, Loc, URL, CS, SGN)$ where:
- $CS = \bigcup_{i=1}^{n} CS_i$.  • $SGN = (GSP, IS)$ where:

  - $GSP = (MS, AS)$ where
    $MS = \text{Mtd. Disc}\{[...](p_1)(p_{n+3})\}, AS = [...]$.
  - $IS = (P, T, W, F, Trc, Tra, L)$ where $P = \{p_i \mid 1 \le i \le n+3\}, T = \{t_i \mid 1 \le i \le n+3\}$
    $W = \{(p_i, t_i) \mid 1 \le i \le n+2\} \cup \{(t_1, p_i), (t_i, p_{n+1}) \mid 2 \le i \le n\} \cup \{(p_{n+1}, t_{n+3}), (t_{n+1}, p_{n+2}), (t_{n+2}, p_{n+3}), (t_{n+3}, p_{n+3})\}, F = \{f(p_1, t_1) = [B], f(p_{n+1}, t_{n+1}) = [B], f(p_{n+1}, t_{n+3}) = [B], ...\}, Trc = trc(t_{n+1}) = [B == true], trc(t_{n+3}) = [B == false], ...\}, \quad Tra = \{tra(t_1) = [B := true], tra(t_{n+1}) = [B := false], ...\},$
    $L = \{(p_i, ISP(S_{i-1})) \mid 2 \le i \le n\} \cup \{(p_1, \tau), (p_{n+1}, \tau), (p_{n+2}, ISP(S_n)), (p_{n+3}, goal)\}$.

Graphically, given $S_1, ..., S_n$, the composite service $(S_1 \square S_2 \square ... \square S_{n-1}) \gg S_n$ is represented by the G-Net shown in Figure 5(a).

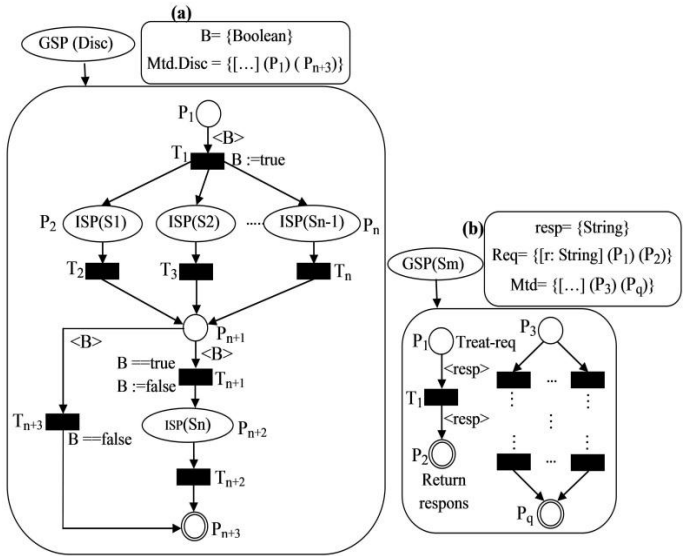
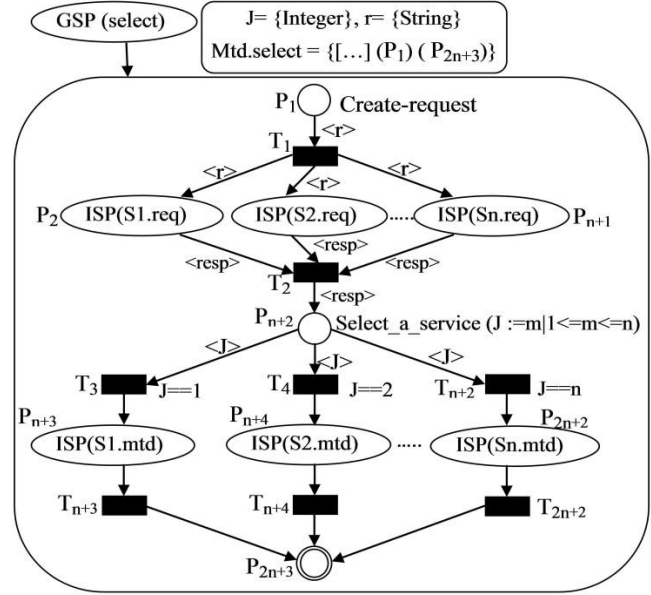

Fig. 5 Discriminator service (a) and service $S_m$ (b)

**Discriminator.** The main goal of the discriminator operator is to increase reliability and delays of the services through the Web. For customers, best services are those which respond in optimal time and are constantly available. The composite construct obtained by applying the Discriminator operator submits redundant orders to different services performing the same task ($S_1, S_2, ..., S_{n-1}$ for example). The first service which responds to the request activates the service $S_n$. All other late responses will be ignored.

Fig. 6 Selection service

**Selection.** Let's have "n" web services ($S_1, ..., S_n$) provide differently the same service, these services are extended by a specific method (req) which receives and answers to certain requests.(An example one of these services is presented in figure 5(b)). Select is an operator which permits, after treating the answers, to choose one service among others which have respond to the same request, this operator offers the possibility to benefit of the best service. The choice is related to several criteria, for example, the best services of sale those which represent seductive prices and suitable delay of delivery.

**Definition 10.** The service $[S_1 | S_2 | ... | S_n]$ is defined as $[S_1 | S_2 | ... | S_n] = (NameS, Desc, Loc, URL, CS, SGN)$ where:

- $CS = \cup_{i=1}^{n} CS_i$.
- $SGN = (GSP, IS)$ where :

  - $GSP = (MS, AS)$ where
    $MS = Mtd.Select\{[...](p_1)(p_{2n+3})\}$,
    $AS = [J: Integer, r: String, ...]$.
  - $IS = (P, T, W, F, Trc, Tra, L)$ where $P = \{p_i \mid 1 \leq i \leq 2n+3\}, T = \{t_i \mid 1 \leq i \leq 2n+2\}$
    $W = \{(p_i, t_2), (p_{n+2}, t_{i+1}), (p_{i+n+1}, t_{i+n+1}), (t_i, p_i), (t_{i+1}, p_{i+n+1}), (t_{i+n+1}, p_{2n+3}) \mid 2 \leq i \leq n+1\} \cup \{(p_1, t_1), (t_2, p_{n+2})\}$, $F = \{f(t_1, p_i) = r, f(p_i, t_2) = [resp] \mid 2 \leq i \leq n+1\} \cup \{f(p_{n+2}, t_i) = [j] \mid 3 \leq i \leq n+2\} \cup \{f(p_1, t_1) = [r], f(t_2, p_{n+2}) = [resp], ...\}, Trc = \{Trc(t_i) = [j == i-2] \mid 3 \leq i \leq n+2\} \cup \{...\}, Tra = \{...\}, L = \{(p_i, ISP(S_{i-1}.req)), (p_{i+n+1}), ISP(S_{i-1}.mtd) \mid 2 \leq i \leq n+1))\} \cup \{(p_1, Create - request), (p_{n+2}, Select - Service), (p_{2n+3}, goal)\}$.

Graphically, given $S_1, ..., S_n$, the composite service $[S_1 | S_2 | ... | S_n]$ is represented by the G-Net shown in Figure 6.

**Refinement.** The refinement permits to replace certain operations of the service by more detailed ones. Refinement is the transformation of a design from a high level abstract form to a lower level more concrete form hence allowing hierarchical modeling.

**Definition 11.** The service Re $Ref(S, a, A)$ is defined as $Ref(S, a, A) = (NameS, Desc, Loc, URL, CS, SGN)$ where:

- $CS = \{CS_1 \cup CA$ if $a \in L_{(S1)}(P_{(S1)})$, $CS_1$ otherwise.
- $SGN = (GSP, IS)$ where :

  - $GSP = (MS, AS)$ where
    $MS = S.Mtd\{[...](p_a)(p_q)\}$ if $L_{(S1)}(p_1) = a$ (where $p_1$ is the initial place of $S_1$ and $p_a$ is the initial place of A), $MS = S.Mtd \{[...](p_1)(p_q)\}$ otherwise. $AS = [...]$.
  - $IS = (P, T, W, F, Trc, Tra, L)$ where $P = P_{(S1)} \backslash L_{(S1)}^{-1}(a) \cup P_A$ if $a \in L_{(S1)}(P_{(S1)})$, $P_{(S1)}$ otherwise. $T = T_{(S1)} \cup T_A$ if $a \in L_{(S1)}(P_{(S1)})$, $T_{(S1)}$ otherwise. $W = \{W_{(S1)} \backslash \{(x, y) \mid x \in L_{(S1)}^{-1}(a) \vee y \in L_{(S1)}^{-1}(a)\} \cup \{(t_i, p_j) \mid \exists p_k \in t_i \bullet \wedge p_k \in L_{(S1)}^{-1}(a) \wedge p_j \in P_A \wedge \bullet p_j = \emptyset\} \cup \{(p_i, t_j) \mid \exists p_k \in \bullet t_j \wedge p_k \in L_{(S1)}^{-1}(a) \wedge p_i \in P_A \wedge p_i \bullet = \emptyset\} \cup W_A$ if $a \in L_{(S1)}(P_{(S1)})$, $W_{(S1)}$ otherwise. $F = \{F_{(S1)} \backslash \{f_{(S1)}(x, y) \mid x \in L_{(S1)}^{-1}(a) \vee y \in L_{(S1)}^{-1}(a)\} \cup \{f(t_i, p_j) = f_{(S1)}(t_i, p_k) \mid p_k \in t_j \bullet \wedge p_k \in L_{(S1)}^{-1}(a) \wedge p_j \in P_A \wedge \bullet p_j = \emptyset\} \cup \{f(p_i, t_j) = f_{(S1)}(p_k, t_j) \mid p_k \in \bullet t_j \wedge p_k \in L_{(S1)}^{-1}(a) \wedge p_i \in P_A \wedge p_i \bullet = \emptyset\} \cup F_A$ if $a \in L_{(S1)}(P_{(S1)})$, $F_{(S1)}$ otherwise.
    $Trc = Trc_{(S1)} \cup Trc_A$ if $a \in L_{(S1)}(P_{(S1)})$, $Trc_{(S1)}$ otherwise.
    $Tra = Tra_{(S1)} \cup Tra_A$ if $a \in L_{(S1)}(P_{(S1)})$, $Tra_{(S1)}$ otherwise. $L = L_{(S1)} \backslash \{(p_i, a) \mid p_i \in P_{(S1)}\} \cup L_A$ if $a \in L_{(S1)}(P_{(S1)})$, $L_{(S1)}$ otherwise.

Figure 7 shows an example of a refined service.

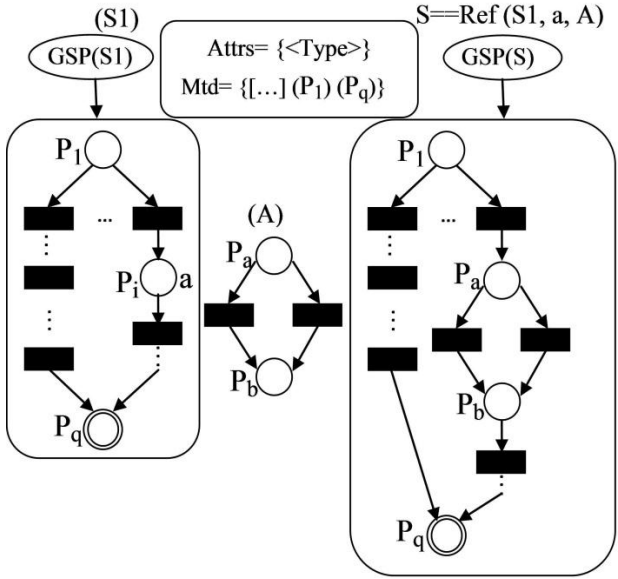

Fig. 7 A refinement example

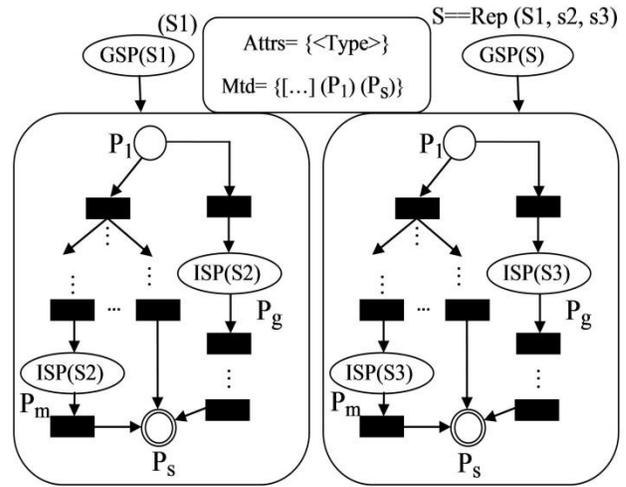

Fig. 8 Service $S_1$ and Rep $(S_1, S_2, S_3)$

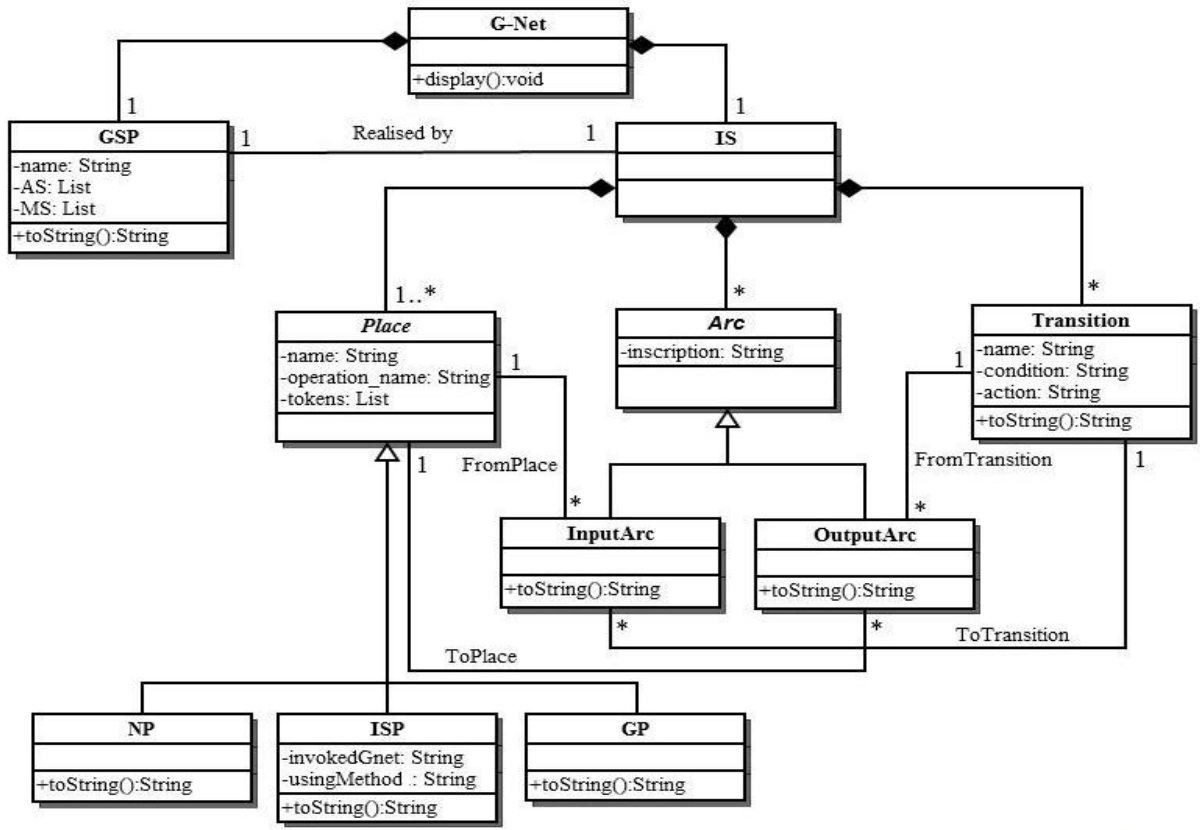

Fig. 9 G-Nets Meta-Model

**Replace.** The operator replace permits to replace a component service by another one inside a composite service. This operation permits to solve the problem of the availability. i.e if a service is not available; it can be replaced by another one, which insures the same functions. Here it is a question of behavior equivalence.

**Definition 12.** The service $Rep(S_1, S_2, S_3)$ is defined as $Rep(S_1, S_2, S_3) = (NameS, Desc, Loc, URL, CS, SGN)$ where:

- $CS = \{CS_1 \setminus CS_2 \cup CS_3 \quad if \ a \in L_{(S1)}(P_{(S1)}),$
  $CS_1$ otherwise.
- $SGN = (GSP, IS)$ where :

  - $GSP = (MS, AS)$ where
    $MS = Mtd.Mtd\{[\ldots](p_1)(p_S)\}, \quad AS = [\ldots].$
  - $IS = (P, T, W, F, Trc, Tra, L)$ where $P = P_{(S1)}$, $T = T_{(S1)}$, $W = W_{(S1)}$, $F = F_{(S1)}$, $Trc = Trc_{(S1)}$, $Tra = Tra_{(S1)}$,
    $L = L_{(S1)} \setminus \{ (p_i, ISP(S_2)) \mid p_i \in P_{(S1)} \} \cup \{(p_i, ISP(S_3)) \mid p_i \in L_{(S1)}^{-1}(ISP(S_2))\} \ if \ CS_2 \subset CS_1$, $L_{(S1)}$ otherwise.

Given $S_1, S_2$ and $S_3$, $Rep(S_1, S_2, S_3)$ is represented by the G-Net shown in Figure 8.

## 4. G-Nets Meta-modeling

In this section, a meta-model for G-nets has been defined, as shown in the figure 9. The meta-formalism, used in this work, is the UML class diagram model. Our meta-model is composed, mainly, of six classes (G-Net, GSP, IS, Place, Transition and Arc).

- **Class G-Net**: it builds the final G-net model from the GSP and the internal structure (IS).
- **Class GSP**: it represents the G-net Generic Switch Place (GSP). It has three attributes. The first one is a key attribute "*name*". However, the two other attributes are lists. The first list, "*AS*", contains G-net variables; and the second list, "*MS*", includes G-net methods.
- **Class IS**: it represents the G-net internal structure. It consists of a set of places and transitions connected by arcs.
- **Class Place**: it represents the G-net place. It has three attributes: "*name*", "*operation-name*", and "*tokens*". It is a three classes super-class: "*NP*" for Normal place, "*GP*" for Goal place, and "*ISP*" for instantiated switch place, the last class contains two attributes: the "*invokedGnet*" and the "*usingMethod*"; which, respectively, represents the G-net name and the method name to be invoked.
- *Class transition*: it represents the G-net transitions. It has three attributes: "*name*", "*condition*", and "*action*".
- *Class arc*: it represents the G-net arcs. It has a "*inscription*" attribute, which represents the arc description. It is a two classes super class: the Input Arc for arcs going from places to transitions, and Output Arc for arcs going from transitions to places.

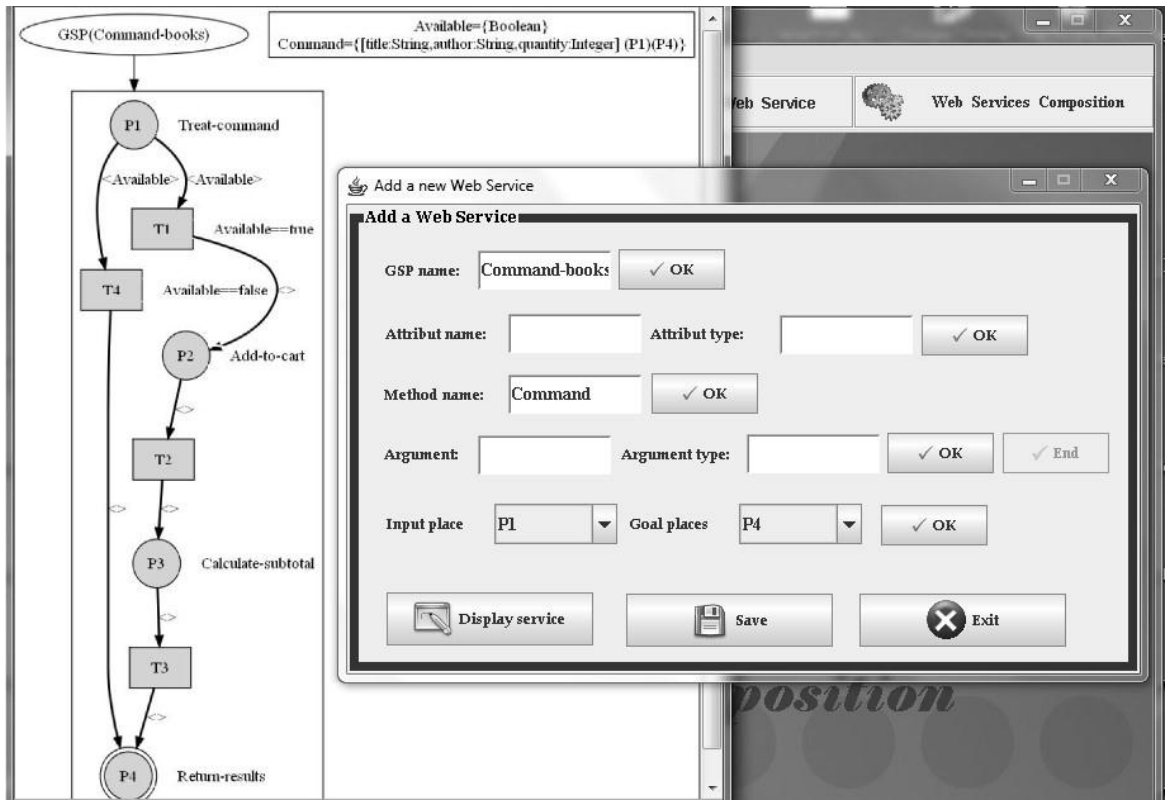

Fig. 10 Service "Command-books"

The associations "*fromplace*", "*toplace*", "*fromtransition*" "*totransition*", assure that G-net arcs do not connect; but places to transition, or vice versa.

## 5. Implementation

In order to automate this approach, a java application has been developed, by us; on the basis of the definite meta-model for G-nets, besides a set of composition rules, that have been formally defined, previously.

Web services specification, back up, modification, and composition, in terms of G-net, are allowed by this application. The main functionality of this application is that of the composition. The latter takes as input a service or a set of services on which it applies an operation or a set of operations to return a composite or a refined service as a result.

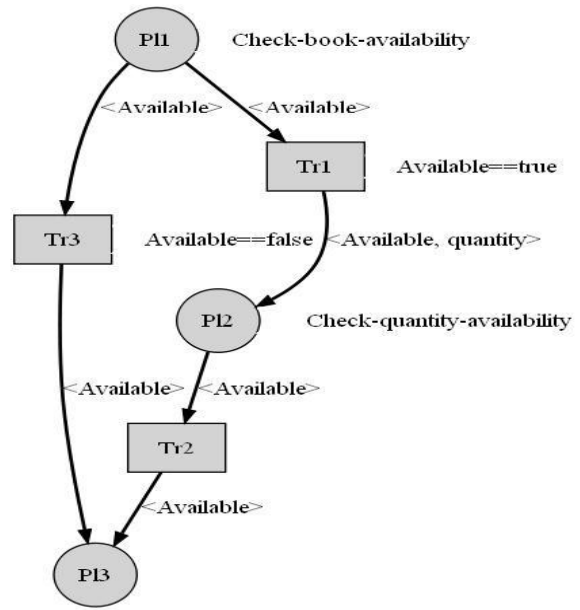

Fig. 11 The block (B)

For more illustration the figure 12 represents the result of the application of a refinement operation on the service *"Command-books"* represented in the figure 10. The refinement affects the place ($p_1$) labeled by the action name *"Treat-Command"*; which is replaced by the well-formed block (B), represented in the figure 11. The resulting service verifies, at first, the book availability. If yes, it verifies if the amount in question exists in the stock. In this case the command is added to the card, therefore the subtotal is calculated.

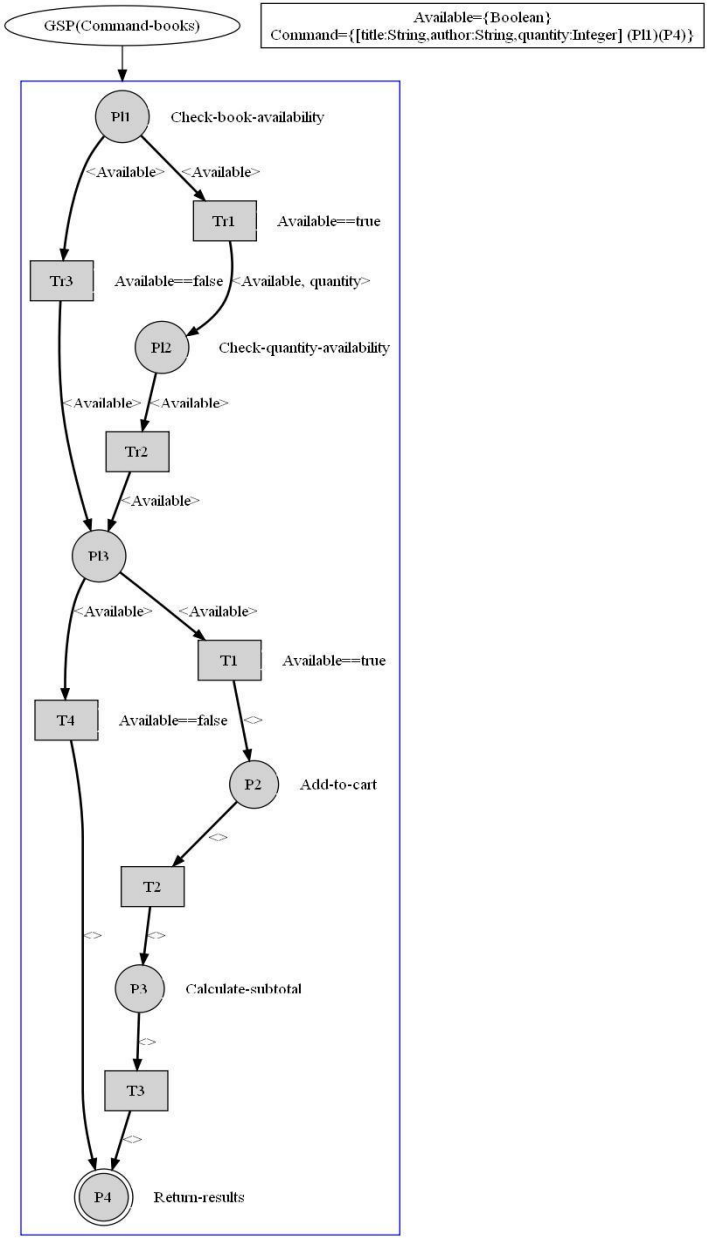

Fig. 12 Service Ref(Command-books, "Treat-command", B)

## 6. Web service verification

The formal verification of the web services composition is an important topic; its interest is to provide correct web services. This is vital for the companies. A web service which contains errors may lead to the clients' loss.
Two approaches can be distinguished, in the formal verification of the web services composition. The first approach focuses on the verification of the composite web services that have already been developed and expressed in languages such as BPEL. In this approach, the composite service code is translated in a formal specification language. The second approach use the formal verification techniques in the development phase, i.e. the web services composition is, directly, specified using a formal description language, before generating the code, which assures the composite web services accuracy.

In [14], the authors have explained that Petri-nets allow the properties verification, each one has a specific meaning in the web services composition field. Among these properties, we cite the reachability, the boundedness study, the dead marking, fairness...etc.

Concerning G –nets, in [15] Dang has presented a method for transforming a G-net model into another equivalent PrT-Nets model. This method has been automated by Kerkouche et al. in [16] while using the graph transformation tool ATOM3 [17]. The purpose of this transformation is to apply the techniques of the formal analysis of PrT-Nets on G-nets specifications. One of the most used analyzers for PrT-Nets is PROD [18], which is a reachability analysis textual tool, the language of describing the PrT-Nets models, in PROD, is C language extended by directives allowing to describe Petri nets. The automatic generation of the PROD description was presented in [16]; the authors have always used the ATOM3 tool to generate a text file, which contains a PROD description from a given PrT-Nets model. This description is compiled by the PROD analyzer to obtain an executable program, which generates the PrT-net model reachability graph.

The reachability analysis offers a precise way to verify and observe the behavior of distributed and concurrent systems. With reachability analysis it is possible to verify some properties of the system and also find out all the chains of events which have caused error states within the system.

Figures 13 and 14, respectively illustrates the result of the transformation of the G-Net model, presented earlier in Figure 12 to the equivalent PrT-Net model and PROD description, generated from the resulting model PrT-Net.

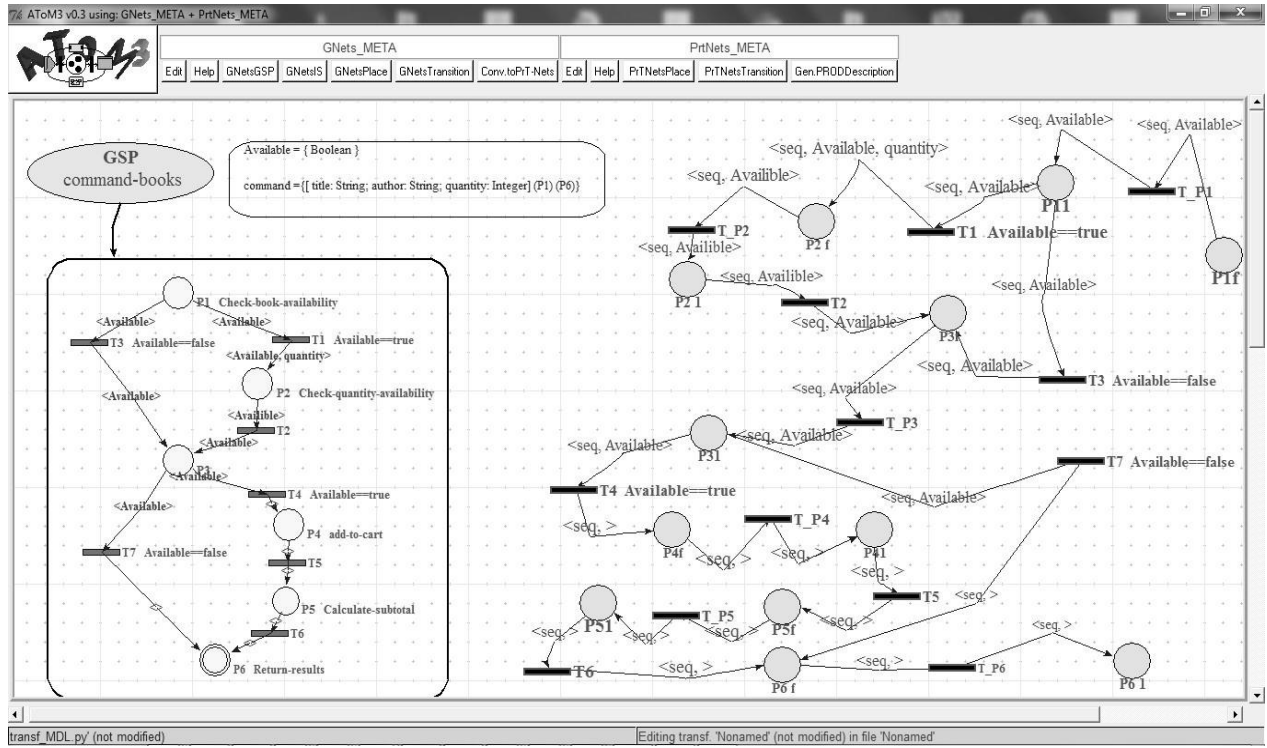

Fig. 13 The generated equivalent PrT-Net model of "Command-books"

Fig. 14 The generated PROD description

# 7. Related Work

The composition of web services is a topic; which has drawn the attention of the researchers, who attempt to offer rigorous semantic models, languages, and platforms for the proposal of effective results for that topic.

Many languages have emerged to attempt to solve this problem. These languages describe the interactions between different web services providers and their customers. WSCL (Web Service Conversation Language) [19] and WSCI (Web Service Choreography Interface) [20] are languages, which are associated with the choreography [21]. Whereas, Orc [22], XPDL [23], and BPEL4WS [24], are associated with the orchestration [25], which tackles the problem in a centralized manner, where web services compositions are carried out by a component, which directs calls to web services.
In parallel, choreography deals with the problem in a distributed manner. Contrary to the orchestration, there is no coordinating service; each service is conscious of the services likely to call it; as well as those, which it must call to execute the business process. BPEL is the language the most used; it is the fusion of XML (Business Process Language) specifications and WSFL (Web Service Flow Language). BPEL contains the characteristics of a block structured language of XLANG, and the characteristics of a directed graph of WSFL.
Unfortunately, it comes to textual and executable languages designed to satisfy the implementation phase of a composite web service, which neglects, in fact, the step of the specification, which is important, because it facilitates the global comprehension of the system, and the development task. Moreover, the formal analysis of the proposed languages is impossible, because of their lack of the formalism.

Approaches of web services composition based on ontologies; such as OWL-S [26], SAWSDL [27] and [28], use consensual terms of predefined ontologies for declaring the pre-conditions, and the effects of the in question services. In the two first works, inputs and outputs are expressed by concepts. While, in [9], they are expressed in terms of instance- based graph models. One of the principal criticisms about OWL-S is that it does not impose any constraint on, so that, the service profile and the service model be coherent [29]. Although, SAWSDL does not require many efforts for the developers used to WSDL, it does not allow the definition of non-functional properties. Nevertheless, besides being very popular, OWL-S has the advantage of being ancient in terms of tooling.

In general , though these approaches have the advantage of the clear understanding of the messages meaning, their main disadvantage is the difficulty of finding the explicit purpose of these services . This latter constitutes a key element, while the composition, by the planners AI [30].

Different formalisms, for modeling and specifying the services composition, have been proposed. We cite: Automata, contracts, UML activity diagrams, process algebras and Petri nets.

Automata are very known models in the field of the formal specification of systems. To model the service composition using an automaton, a state to each service invocation, and an event to each service reply, and a second state to indicate the end of the invocation activity, are, generally, assigned. Several researchers have proposed the use of automata in the field of the web services composition. Among these approaches: the Roman model [31], which is based on FSA (Finite State Automata), the Columbo model [32], which is an extension to the Roman model, which adds the support for data and communication based on the message exchange. In [33], the authors have proposed to model web services using AFA (alternating finite automata). Another approach, called COCOA, has been introduced in [34]; the main idea of COCOA is to convert the OWL-S processes into FSA (Finite State Automata). This allows the conversion of the composition of the services into an automata analysis problem.
All these approaches require an interaction with the developer to provide, beforehand, a detailed specification of the composite service. To solve this problem, authors, in [35], propose the use of DFSA (deterministic finite-state automata), in this approach, the developers only define the exactness constraints on the composition. The behavior of the composite service is automatically synthesized. Mitra et al. have proposed to use input output automata (i/o automata) for modeling the web services composition, in [36].the approach selected in [36], consists of identifying all the feasible compositions using the imported services, and verifying if one of the compositions provides the desired functionality. The disadvantage of this technique is that computation, of the possible compositions, is exponential in relation to the number of the services to compose. Therefore, Mitra et al. have published second article [37] to improve this method by using a logic programming technique.
The disadvantage of using automata remains in the fact that the last mentioned do not allow the direct modeling of the parallel split (AND split) and the synchronization (AND join). This is due to the fact that their semantics does not allow the competition modeling. Although, Yan and Dague have proposed, in [38], a solution for this problem; the idea is to model each parallel execution branch by independent automaton, and define synchronization events to realize their interconnection; but, unfortunately, this requires the input and the output state duplication, in each parallel branch.

The approach [39] consists in using the notion of contracts. The last mentioned are graph transformation rules. They are specified by the assertions expressing the rights and duties of the providers or the customers. These assertions can be pre-conditions, post conditions, or invariants.
The advantage of the graph transformation rules is represented in their ability to bring advanced operational interpretations, which cannot be expressed with simple logical expressions. However, they present disadvantage in the case of the complex composition. This approach remains inadequate, if we want to make a dynamic or a semi- automatic services composition.

Other approaches have used the UML activity diagrams, for compositing web services. To model the web services composition, using the last mentioned, a service execution is, generally, represented by an action; the transition to this action models the call to a service, and the output indicates the end of the service execution.

In [40], the authors present an approach for the business process modeling using the activity diagram they have, also discussed the transformation of the results obtained in BPEL 1.1. This transformation was implemented and updated to support UML 2.0 in [41], but the disadvantage of this implementation is, in fact, that it is, always, based on BPL, 1.1.other authors have presented a work based on the UML 1.4 utilization for the development of the composite web service following MDA principles [42].
The same idea has been used in [43]; this work is based on UML 2.0. However, the authors have pointed out that the model is not expressive enough. Besides, it is worth to note that UML is a semi-formal language.

Process algebras are a mathematic formalism for describing and studying the concurrent systems. Different research papers have used these formalisms for web service composition. We cite as an example, the work of J.Camara et al., who have proposed a formalization technique based on CCS, for the choreography of the web service in [44]. Liu et al. have used the CCS algebra for modeling and specifying web services to reason on the composition behavioral properties. In [45], semantics of the orchestration language BPEL is, this time, specified by using π-calculus. Nevertheless, this work does not deal with some parts of BPEL, such as data management; and this is not surprising, because π-calculus does not allow data manipulation.

Although process algebras are well adapted for the complex systems description, their textual notation makes them less readable than Petri-nets.

Concerning Petri-nets, Onyang et al. have presented a complete translation of BPEL in Petri-nets in [46]. Besides, they have also demonstrated the ability of the last mentioned in modeling the web services composition. In the same vein of [12], our approach realizes the web service composition, using Petri-nets based algebra. Their model is expressive enough; but data types cannot be distinguished, because they use elementary Petri-nets. Instead of the last mentioned, our model uses a high level Petri-nets type, called G-nets. In this formalism, the arrows can be labeled by constants or variables specifying the token parameters. Like arcs, conditions can be associated with transitions. The value of the parameters for which the transition is firable is precised by these conditions. Consequently, data type can be distinguished in our proposal. The work of [47], also, deals with this problem by modeling web services and their composition via colored Petri-nets [48], in relation to [47]. The main advantage of our approach is that the model requires great efforts, when modeling complex services and produces smaller models. Similarly to our approach, the proposal of [49] is also based on process modeling of web services composition by a kind of object-oriented Petri-nets. However, our approach is formally defined. Moreover, it is based on a well-founded framework, namely G-nets.
It should be noted that all these works are purely theoretical, and there is no work among the last mentioned that has automated its approach.

## 8. Conclusion

In this paper, a simple efficient approach for Web services composition has been proposed. Benefiting from the formal, modular, and object-oriented aspects of G-Nets is the main advantage of this approach. Besides, the specification and prototyping of complex Web services are allowed by modeling using G-Nets. A G-Net based algebra is developed for Web services composition. The formal semantics of the composition operators is defined, by means of G-Nets, in this context. Using this underlying framework provides a rigorous approach for verifying properties and detecting inconsistencies between Web services.
In order to automate this approach, we have developed a tool based on the meta-model that we have proposed to assure the conformity of the used models to the latter, and on the formally defined composition rules set.
Java, the programming language; has been chosen, because of its compatibility with UML object-oriented concepts, on the one hand. And of its various characteristics: simplicity, robustness, portability, and security, on the other hand. Java is a standard and a dynamic language. The development we have carried out allows a kind of suppleness for possible changes, or improvement; in the meta-model, or in the set of the composition rules by changing the corresponding classes. Our approach has been illustrated with an example.

In a future, this approach is planned to be extended with advanced operators, which can support more complex combination of Web services. We can also use the well-known reduction technique on the obtained models before performing the analysis and the verification to optimize the models. An other perspective, in our work, is to propose a M2T transformation approach, which allows transforming G-nets models to Maude specifications, to use the Maude system as a means for analyzing and simulating the systems modeled by G-nets.